\documentclass{article}

\usepackage[margin=3cm]{geometry}

\input babarsym

\newcommand{\romegarhop}{\ensuremath{15.9\pm1.6\pm1.4}} % BF measurement
\newcommand{\somegarhop}{\ensuremath{9.8}}		% significance
\newcommand{\Aomegarhop}{\ensuremath{-0.20\pm0.09\pm0.02}}  % Ach
\newcommand{\fLomegarhop}{\ensuremath{0.90\pm0.05\pm0.03}}  % polarization

\newcommand{\romegarhoz}{\ensuremath{0.8\pm0.5\pm0.2}} % BF measurement
\newcommand{\ulomegarhoz}{\ensuremath{1.6}}	% 90% CL UL

\newcommand{\romegafz}{\ensuremath{1.0\pm0.3\pm0.1}} % BF measurement
\newcommand{\ulomegafz}{\ensuremath{1.5}}	% 90% CL UL

\newcommand{\romegaKstz}{\ensuremath{2.2\pm0.6\pm0.2}} % BF measurement
\newcommand{\somegaKstz}{\ensuremath{4.1}}		% significance
\newcommand{\AomegaKstz}{\ensuremath{0.45\pm0.25\pm0.02}}  % Ach
\newcommand{\fLomegaKstz}{\ensuremath{0.72\pm0.14\pm0.02}} % polarization

\newcommand{\romegaKpiSwavez}{\ensuremath{18.4\pm1.8\pm1.7}}

\newcommand{\somegaKpiSwavez}{\ensuremath{9.8}}	% significance

\newcommand{\AomegaKpiSwavez}{\ensuremath{-0.07\pm0.09\pm0.02}}

\newcommand{\romegaKtstz}{\ensuremath{10.1\pm2.0\pm1.1}}
\newcommand{\somegaKtstz}{\ensuremath{5.0}}		% significance
\newcommand{\AomegaKtstz}{\ensuremath{-0.37\pm0.17\pm0.02}}
\newcommand{\fLomegaKtstz}{\ensuremath{0.45\pm0.12\pm0.02}} % polarization

\newcommand{\romegaKstp}{\ensuremath{2.4\pm1.0\pm0.2}} % BF measurement
\newcommand{\somegaKstp}{\ensuremath{2.5}}		% significance
\providecommand{\ulomegaKstp}{\ensuremath{3.8}}

\newcommand{\AomegaKstp}{\ensuremath{0.29\pm0.35\pm0.02}}	% Ach
\newcommand{\fLomegaKstp}{\ensuremath{0.41\pm0.18\pm0.05}}      % polarization

\newcommand{\romegaKpiSwavep}{\ensuremath{27.5\pm3.0\pm2.6}} % BF measurement
\newcommand{\somegaKpiSwavep}{\ensuremath{9.2}}		% significance

\newcommand{\AomegaKpiSwavep}{\ensuremath{-0.10\pm0.09\pm0.02}}

\newcommand{\romegaKtstp}{\ensuremath{21.5\pm3.6\pm2.4}} % BF measurement
\newcommand{\somegaKtstp}{\ensuremath{6.1}}		% significance
\newcommand{\AomegaKtstp}{\ensuremath{0.14\pm0.15\pm0.02}}
\newcommand{\fLomegaKtstp}{\ensuremath{0.56\pm0.10\pm0.04}}      % polarization

\newcommand{\rbmrhop}{\ensuremath{-1.8 \pm 0.5 \pm 1.0}}
\newcommand{\sbmrhop}{\ensuremath{-}}

\newcommand{\ulbmrhop}{\ensuremath{1.4}\xspace}

\newcommand{\rbprhoz}{\ensuremath{\msp 1.5\pm 1.5 \pm 2.2}}
\newcommand{\sbprhoz}{\ensuremath{0.4}}
\newcommand{\ulbprhoz}{\ensuremath{5.2}\xspace}

\newcommand{\rbzrhop}{\ensuremath{-3.0\pm 0.9 \pm 1.8}}
\newcommand{\sbzrhop}{\ensuremath{-}}
\newcommand{\ulbzrhop}{\ensuremath{3.3}\xspace}

\newcommand{\rbzrhoz}{\ensuremath{-1.1\pm 1.7^{+1.4}_{-0.9}}}
\newcommand{\sbzrhoz}{\ensuremath{-}}
\newcommand{\ulbzrhoz}{\ensuremath{3.4}\xspace}

\newcommand{\rbpKstz}{\ensuremath{\msp 2.9\pm 1.5 \pm 1.5}}
\newcommand{\sbpKstz}{\ensuremath{1.5}}
\newcommand{\ulbpKstz}{\ensuremath{5.9}\xspace}

\newcommand{\rbzKstz}{\ensuremath{\msp 4.8\pm 1.9^{+1.5}_{-2.2}}}
\newcommand{\sbzKstz}{\ensuremath{2.0}}
\newcommand{\ulbzKstz}{\ensuremath{8.0}\xspace}

\newcommand{\rbmKstp}{\ensuremath{2.4^{+1.5}_{-1.3} \pm 1.0}}
\newcommand{\sbmKstp}{\ensuremath{1.7}}
\newcommand{\ulbmKstp}{\ensuremath{5.0}\xspace}

\newcommand{\rbzKstp}{\ensuremath{0.4^{+2.0+3.0}_{-1.5-2.6}}}
\newcommand{\sbzKstp}{\ensuremath{0.1}}
\newcommand{\ulbzKstp}{\ensuremath{6.7}\xspace}

\newcommand{\calB}{\ensuremath{{\cal B}}}
\newcommand\dbline{\noalign{\vskip 0.10truecm\hrule}\noalign{\vskip 2pt}\noalign{\hrule\vskip 0.10truecm}}
\newcommand{\msp}{\ensuremath{\phantom{-}}}
\newcommand\etal{{\it et al.}}
\newcommand{\jhep}[1]{{\it JHEP}\ #1}

\title{Branching fractions and charge asymmetries in charmless
  hadronic $B$ decays at \babar} 

\author{Pietro Biassoni$^{\dagger}$\\ (On behalf of the \babar\
  Collaboration)\\ {\small \emph{$^{\dagger}$ Universit\`a degli Studi and
    INFN Milano, via Celoria 16, I-20133 Milano, Italy.}}}
\date{}

\begin{document}

\maketitle

\begin{abstract}
We present measurements of branching fraction,
polarization and charge asymmetry in charmless hadronic
$B$ decays with $\eta$, \etapr, $\omega$, and $b_1$ in the
final state. All the results use the final \babar\ dataset.
\end{abstract}

\section{Introduction}
Experimental measurements of branching fraction, polarization and
\CP-violating charge asymmetries in rare $B$ decays are important tests of the
Standard Model (SM) and its extensions. 
Several predictions are available for these quantities, using
different theoretical approaches~\cite{theor,BN}.
All these quantities may provide sensitivity to the presence of heavy
non-SM particles in the loop diagrams.   

The large branching fraction difference between $\eta^{\prime} K$ and
$\eta K$ seems to be explained in the SM contest~\cite{Lipkin}.
Rates of the decay modes to $\eta\eta$, $\eta\phi$, \etapr\etapr, and
$\etapr\phi$ are used in flavor SU(3)-based
calculations~\cite{BN,Gross},
to constraint the unsigned difference between the \CP-violating
parameter $S$ measured in $\etapr K^0$ and $\phi K^0$ and $\sin2\beta$ measured in $J/\psi K^0$.
The charge asymmetry ${\cal A}_{ch}$ is expected to be sizable in $\eta K^+$ and
suppressed in $\etapr K^+$ decays~\cite{BN}.

In $B\to VV$ decays (where $V$ is a vector), simple helicity arguments
predict a longitudinal polarization fraction $f_L$ close to 1. 
In 2003 both \babar\ and Belle measured  $f_L \sim 0.5$ in 
$B\to\phi K^*(892)$~\cite{phiKst}. Possible explanations for 
this puzzle have been proposed within the SM 
\cite{VVBSMrefs} and in new physics scenarios \cite{nSMetc}.

\section{Analysis Technique}
Results shown in this paper are based on a sample of
$465\times10^6$ \BB\ pairs collected at a center-of-mass energy $\sqrt{s}$
equal to the mass of the \FourS\ resonance at the PEP-II asymmetric
\epem\ collider, at the SLAC National Accelerator Laboratory, and
recorded by the \babar\ detector~\cite{BABARNIM}. 

$B$ meson is reconstructed into 
$\eta\pip$, $\eta K$, $\eta\eta$, $\eta\omega$, $\eta\phi$,
$\eta^{\prime}\pip$, $\eta^{\prime} K$, $\eta^{\prime}\eta^{\prime}$,
$\eta^{\prime}\omega$, $\eta^{\prime}\phi$, $\omega K^*$, $\omega
f_0(600)$, $\omega\rho$, $b_1 K^*(892)$, and $b_1\rho$ final states.
In $\omega K^*$, we consider either $K^*(892)$, $(K\pi)_0$, and
$K^*_2(1430)$. The $B$ meson is kinematically characterized by 
$\Delta E\equiv E_B-\frac{1}{2}\sqrt{s}$ and $\mes \equiv \sqrt{s/4 -
\vec{p}_B^2}$, where $(E_B,\vec{p}_B)$ is the $B$ meson four-momentum vector
expressed in \FourS\ rest frame.

Background arises primarily from random combinations of particles in
$\epem\to\qqbar$ events ($q=u,d,s,c$).
We suppress this background with requirements on event shape variables
and on the energy, invariant mass and particle
identification signature of the decay products.
For $VV$, and vector-tensor $VT$ decays, we define the helicity
angles $\theta_1$ and $\theta_2$, where the subscript refers to $B$
daughters. For two (three) body decay, $\theta_i$ is defined as the angle between the
direction of the recoiling $B$ and the direction of one of the
resonance daughters (the normal to the plane identified by the daughter
decay products).

For each mode, results are obtained from extended maximum likelihood
fits with input variables $\Delta 
E$, \mes, and the output of a Fisher discriminant that combines
different event shapes variables.  
Where useful, the masses of $B$ daughters are included in the fit.
In $\omega K^*$ and $\omega \rho$, $f_L$ and $f_T = 1-f_L$ are extracted using the knowledge of the decay angular distribution:
\begin{eqnarray} &
  \frac{{\mbox{ d$\Gamma$}}}{\mbox{ d$\cos\theta_1$d$\cos\theta2$}}
=  \left\{ \begin{array}{l} f_T\sin^2\theta_1\sin^2\theta_2 +
                            4f_L\cos^2\theta_1\cos^2\theta_2
                            \;\;\;\;\;\;\;\;\;\;\;\;\;\;\;\;\;\;
                            \;\;\;\;\;\;\;\;\;\;\;
                            \mbox{for } B\rightarrow VV \\
                           f_T\sin^2\theta_1\sin^2\theta_2\cos^2\theta_2
                           +\frac{f_L}{3}\cos^2\theta_1(3\cos^2\theta2
                           -1)^2 \;\;\;\;\;\;\;\mbox{for } B\rightarrow VT \\
         \end{array} 
  \right.
\end{eqnarray}
\section{Results}
In Table~\ref{tab:res} we report the branching fraction $\calB$ and the $\calB$ upper limit (UL) at 90\% confidence level (CL), the significance 
$S$ (with systematic uncertainties included), the charge asymmetry ${\cal A}_{ch}$,
and $f_L$, for each decay mode~\cite{currRes}. 
\begin{table}[!h]
\begin{center}
\small
\caption{Results for modes presented in this paper .}
\label{tab:res}
\begin{tabular}{l|c|c|c|c|c}
\dbline
Decay & $\calB$ ($10^{-6}$) & $\calB$ UL ($10^{-6}$) & $S$ ($\sigma$)&
$A_{ch}$ & $f_L$\\
Mode & & & & \\ 
\dbline
$\eta\pip$ & $4.00 \pm 0.40 \pm 0.24$ & -- & -- & $-0.03\pm 0.09
\pm0.03$ & --\\
$\eta K^0$ & $1.15^{+0.43}_{0.38}\pm0.09$ & 1.8 & 3.5 &  --&--\\
$\eta K^+$ & $2.94^{+0.39}_{-0.34} \pm 0.21 $ & -- & -- & $-0.36 \pm
0.11 \pm 0.03$ & --\\
$\eta\eta$ & $0.5 \pm 0.3\pm 0.1$ &  1.0 & 1.9 & -- & --\\
$\eta\omega$ & $0.94^{+0.35}_{-0.30}\pm 0.09$ & 1.4 &3.7 & -- & --\\
$\eta\phi$ & $0.2\pm 0.2\pm 0.1$  & 0.5 & 1.4 & -- & -- \\
$\eta^{\prime}\pip$ & $3.5 \pm 0.6 \pm0.2$ & -- & -- & $+0.03\pm
0.17 \pm 0.02 $ & --\\
$\eta^{\prime} K^0$ & $68.5 \pm2.2 \pm 3.1$& --& --&--& --\\
$\eta^{\prime} K^+$ & $71.5 \pm 1.3 \pm 3.2$ & -- & -- &
$+0.008^{+0.017}_{-0.018} \pm 0.009 $ & --\\
$\eta^{\prime}\eta^{\prime}$ & $0.6^{+0.5}_{-0.4}\pm 0.4$ &1.7 &1.0 &--&--\\
$\eta^{\prime}\omega$ & $1.01^{+0.46}_{-0.38} \pm 0.09$& 1.8 &3.6 & -- & --\\
$\eta^{\prime}\phi$ &  $0.2 \pm0.2\pm 0.3$ & 1.1 & 0.5 & -- & --\\
\hline
$\omega K^*(892)^0$   & \romegaKstz &$-$ & \somegaKstz & +\AomegaKstz
&\fLomegaKstz\\ 
$\omega K^*(892)^+$    & \romegaKstp &\ulomegaKstp &\somegaKstp
&+\AomegaKstp &\fLomegaKstp \\ 

$\omega (K\pi)_0^{*0}$ &\romegaKpiSwavez & -- & \somegaKpiSwavez
&\AomegaKpiSwavez & --\\ 
$\omega (K\pi)_0^{*+}$ & \romegaKpiSwavep & -- & \somegaKpiSwavep &
\AomegaKpiSwavep & --\\ 
$\omega K_2(1430)^{*0}$ & \romegaKtstz & -- &\somegaKtstz & \AomegaKtstz
&\fLomegaKtstz \\ 
$\omega K_2(1430)^{*+}$ & \romegaKtstp& -- & \somegaKtstp & +\AomegaKtstp
&\fLomegaKtstp \\ 
$\omega f_0$ & \romegafz   &\ulomegafz& 4.5 & -- & -- \\
$\omega\rho^0$ & \romegarhoz &\ulomegarhoz &1.9 & -- &0.8 fixed\\
$\omega\rho^+$ & \romegarhop &-- &\somegarhop & \Aomegarhop  &
\fLomegarhop \\
\hline
$b_1^0\rho^0$ & \rbzrhoz  & \ulbzrhoz\ & \sbzrhoz & -- & --\\
$b_1^-\rho^+$ & \rbmrhop  & \ulbmrhop\ & \sbmrhop & -- & --\\
$b_1^0\rho^+$ & \rbzrhop  & \ulbzrhop\ & \sbzrhop & -- & --\\
$b_1^+\rho^0$ & \rbprhoz  & \ulbprhoz\ & \sbprhoz & -- & --\\
$b_1^0 K^*(892)^0$& \rbzKstz  & \ulbzKstz\ & \sbzKstz & -- & --\\
$b_1^- K^*(892)^+$& \rbmKstp  & \ulbmKstp\ & \sbmKstp & -- & --\\
$b_1^0 K^*(892)^+$& \rbzKstp  & \ulbzKstp\ & \sbzKstp & -- & --\\
$b_1^+ K^*(892)^0$& \rbpKstz  & \ulbpKstz\ & \sbpKstz & -- & --\\
\dbline
\end{tabular}
\end{center}
\end{table}
The first error is statistical and  second systematic.
Results for modes containing $\eta$ or \etapr\ meson in the final
states are preliminary.
Significance is taken as $\sqrt{-2\ln\calL_{max}/\calL_0}$,
where $\calL_{max}$ ($\calL_0$) is value of the likelihood at its
maximum (for zero  signal). If
the significance is smaller than $5\sigma$, we calculate a Bayesian UL at 90\% CL,
integrating the likelihood in the positive branching fraction region.
For the well established decay modes $\eta K^+$, $\eta^{\prime}K^0$, and $\eta^{(\prime)}\pip$ we do not report the significance.
In $\omega K^*(892)^+$ with $K^*(892)^+\to\KS\pip$, $f_L$ is fixed to
$0.5$ in the fit.
Main contributions of systematic uncertainties to branching fraction come from fit bias
and uncertainties in the probability density functions parameterization.  The $B\to\etapr K$ decay mode is
systematic limited due to the uncertainties on daughter
branching fractions. 

\section{Conclusions}
We reported measurements for several charmless hadronic $B$
decays.
In $B\to\eta K^+$ we find evidence of direct \CP\ violation at
$3.3\sigma$ level.
$B\to\omega (K\pi)_0^*$  and $B\to\omega K_2^*(1430)$ decays are
observed for the first time. $f_L$ in $\Bp\to\omega K^*(892)^+$ and
$\Bp\to\omega\rho^+$ is consistent with $0.5$ and $1$, respectively,
as expected by theoretical predictions~\cite{VVBSMrefs}.
$f_L$ in $B\to\omega K_2^*(1430)$ is consistent with $0.5$ in
disagreement with $f_L(\phi K^*_2(1430))\sim1$~\cite{phiKstNew}. No
theoretical predictions are available for these modes.
Results in $B\to b_1\rho$ and $B\to b_1 K^*$ are in disagreement with
and seem to be systematically lower than theoretical predictions~\cite{theor}.

\section{Acknowledgements}
I would like to thank all my \babar\ collaborators and in particular
Fernando Palombo, Adrian Bevan and Jim Smith for their support.


\begin{thebibliography}{99}
\bibitem{theor}
Y.~H.~Chen \etal, \jprd{60}, 094014 (1999) [hep-ph/9903453];
C.~S.~Kim \etal, \jprd{67}, 014002 (2003) [hep-ph/0205263];
C.~W.~Chiang \etal, \jprd{69}, 034001 (2004) [hep-ph/0307395];
Z.~Xiao \etal,  \jprd{75}, 014018 (2007) [hep-ph/0607219];
H.-Y.~Cheng \etal, \jprd{77}, 014034 (2007) [hep-ph/0705.3079];
H.-Y.~Cheng and K.-C.~Yang, \jprd{78}, 094001 (2008) [hep-ph/0805.0329].

\bibitem{BN}
M.~Beneke and M.~Neubert, \npb{675}, 333 (2003) [hep-ph/0308039];
C.~W.~Chiang \etal, \jprd{70}, 034020 (2004) [hep-ph/0404073]. 

\bibitem{Lipkin}
H.~J.~Lipkin, \plb{254}, 247 (1991); M.~Beneke and M.~Neubert,
\npb{651}, 225 (2003) [hep-ph/0210085].

\bibitem{Gross}
Y.~ Grossman \etal, \jprd{68}, 015004 (2003) [hep-ph/0303171]; 
M. Gronau \etal, \plb{596}, 107 (2004) [hep-ph/0403287].

\bibitem{phiKst}
\babar\ Collaboration, B. Aubert \etal, \jprl{91}, 171802 (2003) [hep-ex/0307026];\\
Belle Collaboration, K.F.~Chen \etal, \jprl{91}, 201801 (2003) [hep-ex/0307014].

\bibitem{VVBSMrefs}
For a review of SM understandings of the ``{\it polarization puzzle}''
see A.~Datta \etal, \jprd{76}, 034015 (2007) [hep-ph/0705.3915], and
references therein. 

\bibitem{nSMetc}
E.~Alvarez \etal, \jprd{70}, 115014 (2004) [hep-ph/0410096];
P.~K.~Das and K.~C.~Yang, \jprd{71}, 094002 (2005) [hep-ph/0412313];
C.-H.~Chen and C.-Q.~Geng, \jprd{71}, 115004 (2005) [hep-ph/0504145]; 
A.~K.~Giri and R.~Mohanta, \epjc{44}, 249 (2005) [hep-ph/0412107];  %
S.~Baek \etal, \jprd{72}, 094008 (2005) [hep-ph/0508149];
W.~Zou and Z. Xiao, \jprd{72}, 094026 (2005)[hep-ph/0507122];%
Q.~Chang, X.-Q.~Li, and Y.~D.~Yang, \jhep{0706}, 038 (2007) [hep-ph/05610280].

\bibitem{BABARNIM}
\babar\ Collaboration, B.\ Aubert \etal, \nima{479}, 1 (2002) [hep-ex/0105044].

\bibitem{currRes}
\babar\ Collaboration, B.\ Aubert \etal, \jprd{79}, 052005 (2009)
[hep-ex/0901.3703]; hep-ex/0907.1743, submitted to \jprd{};
\jprd{80}, 501101 (2009) [hep-ex/0907.3485].

\bibitem{phiKstNew}
\babar\ Collaboration, B. Aubert \etal, \jprd{78}, 092008 (2008)
[hep-ex/0808.3586].

\end{thebibliography}
\end{document}